\documentclass[11pt,a4paper]{article}

\usepackage[margin=1in]{geometry}
\usepackage{times}
\usepackage{amsmath}
\usepackage{amssymb}
\usepackage{graphicx}
\usepackage{booktabs}
\usepackage{hyperref}
\usepackage{url}
\usepackage{cite}
\usepackage{float}
\usepackage{array}
\usepackage{lineno}
\usepackage{setspace}
\usepackage{titlesec}
\usepackage{abstract}
\usepackage{xurl}
\usepackage{hyperref}
\usepackage{amsmath}
\usepackage{amssymb}
\usepackage{amsthm}

\usepackage{graphicx}
\usepackage{booktabs}
\usepackage{multirow}
\usepackage{xcolor}
\usepackage{algorithm}
\usepackage{algpseudocode}
\hypersetup{
    colorlinks=true,
    linkcolor=blue,
    citecolor=blue,
    urlcolor=blue
}

\titlespacing*{\section}{0pt}{12pt}{6pt}
\titlespacing*{\subsection}{0pt}{10pt}{4pt}

\title{
    \Large\textbf{
    Systematic Evaluation of Adversarial Jailbreaking\\
Attacks and Defenses Across Seven Large Language Models
    }
}

\author{
\begin{tabular}{c}
\textbf{Md. Robiul Islam Niloy}\\
Department of Computer Science and Engineering\\
BRAC University, Dhaka, Bangladesh\\
\texttt{md.niloy26643@gmail.com}
\\[10pt]
\textbf{Abu Sayed Md. Mostafizur Rahaman}\\
Department of Computer Science and Engineering\\
Jahangirnagar University, Dhaka, Bangladesh\\
\texttt{asmmr@juniv.edu}
\end{tabular}
}

\date{November 2026}

\begin{document}

\maketitle

\begin{abstract}
Large language models (LLMs) deployed in safety-critical settings
must resist adversarial inputs designed to elicit harmful outputs.
We present a systematic empirical evaluation of five adversarial
jailbreaking attack families---direct requests, roleplay-based
jailbreaking, hypothetical framing, instruction override, and
obfuscation---across seven open-weight LLMs served locally:
Llama-3-8B, Llama-3.1-8B, Llama-3.2-3B, Mistral-7B,
Qwen2.5-7B, Gemma2-9B, and Phi-4-14B.
We further evaluate four defense mechanisms---keyword blacklisting,
self-reminder prompting, paraphrase-based sanitization, and
perplexity-based filtering---against two high-ASR models (Mistral-7B and Llama-3.1-8B) on a
20-prompt subset.
Across 250 prompts per model, a content judge counted obfuscation attacks
as harmful (HARMFUL or PARTIAL responses) at a pooled rate of 62.3\%, the
highest of any family, although only 7.1\% of responses were rated
HARMFUL. Hypothetical framing was the highest family for three of seven
models. Obfuscation succeeded at nonzero rates on all seven. The content
judge has not yet been validated against human labels, and a preliminary
check by a non-expert labeller found much lower harm rates, so the headline
should be treated as provisional. We find that model size does not significantly predict adversarial
robustness (Spearman $\rho = -0.64$, $p = 0.12$, $n = 7$). A rule-based
judge suggested a large gap between the two Llama-3 generations (16.8\%
and 40.8\%); the content judge does not (17.2\% and 18.4\%). Self-reminder
prompting gave the largest reduction, cutting ASR from
50\% to 10\% on Mistral-7B and from 45\% to 5\% on Llama-3.1-8B
(40 percentage points on each model, against matched no-defense runs,
scored with the rule-based judge; these defense results are provisional
because that judge has known defects). Our
code and prompts are publicly available at
\url{https://github.com/Mdniloykhan/jailbreak-eval-7llms}; model responses
and judge outputs are available on request.

\noindent  \textbf{Keywords:} adversarial robustness, jailbreaking, large language models, AI safety, red teaming, defense mechanisms, obfuscation attacks, LLM evaluation
\end{abstract}
\section{Introduction}
\label{sec:intro}

Large language models have achieved remarkable performance across
a broad range of natural language tasks, and are increasingly
deployed in applications requiring reliable, safety-aligned
behavior~\cite{ouyang2022training,bai2022constitutional}.
Despite advances in alignment training---including reinforcement
learning from human feedback (RLHF) and constitutional AI---these
models remain susceptible to adversarial inputs that circumvent
their safety mechanisms, a class of attacks known as
\emph{jailbreaking}~\cite{perez2022ignore,zou2023universal}.

Jailbreaking attacks exploit the tension between helpfulness
and harmlessness in aligned LLMs. An attacker frames a harmful
request using role-play scenarios, hypothetical settings, encoded
or obfuscated text, or direct instruction override, causing the
model to produce outputs it would ordinarily refuse. The security
implications are significant: successful jailbreaks have been
demonstrated to extract instructions for dangerous activities,
generate disinformation, and bypass content moderation
policies~\cite{wei2023jailbroken,chao2023jailbreaking}.

Existing evaluations are limited in four key ways. First, most
studies evaluate a single model or a small number of
architecturally similar models, limiting generalizability.
Second, few studies systematically compare multiple attack
families under controlled conditions. Third, the relationship
between model characteristics---such as parameter count and
training methodology---and adversarial robustness remains poorly
understood. Finally, the comparative effectiveness of practical
defense mechanisms across different model families has not been
rigorously characterized.

This paper addresses these gaps with the following contributions:

\begin{enumerate}
  \item We systematically evaluate five adversarial attack
  families across seven LLMs spanning five model developers
  (Meta, Mistral AI, Alibaba, Google, and Microsoft),
  covering 3B to 14B parameters.

  \item We show that obfuscation-based attacks have the highest pooled
  content-judged rate of the families tested, with nonzero rates on every
  evaluated model.

  \item We show that model size does not predict adversarial
  robustness, and that the ranking of models depends on how success is measured, so
  ranking claims require a validated judge.

  \item We evaluate four practical defense mechanisms and
  identify self-reminder prompting as the most effective defense in
  our tests on two models (provisional, rule-based judge), with no
  consistent latency overhead observed.

  \item We show that defense effectiveness is strongly
  model-dependent: the paraphrase defense reduced ASR by
  40 percentage points on Llama-3.1-8B (45\% to 5\%) but did not reduce
  it on Mistral-7B (60\% against a 50\% baseline; provisional, rule-based judge), motivating
  per-model defense evaluation before deployment.
\end{enumerate}

\section{Related Work}
\label{sec:related}

\subsection{Jailbreaking Attacks}

Early jailbreaking work relied on manually crafted prompts to
bypass safety filters~\cite{perez2022ignore}. Zou \emph{et al.}
introduced GCG, a gradient-based white-box optimization attack
that appends adversarial suffixes to harmful
queries~\cite{zou2023universal}. PAIR~\cite{chao2023jailbreaking}
uses an attacker LLM to iteratively refine jailbreak prompts in
a black-box setting. Wei \emph{et al.}~\cite{wei2023jailbroken}
identified competing objectives and generalization failures as
root causes of jailbreak susceptibility. Liu
\emph{et al.}~\cite{liu2023autodan} proposed AutoDAN, a
stealthy optimization-based jailbreak that generates fluent
adversarial prompts.

Most directly related to our work is Niloy \emph{et al.}
\cite{niloy2025jailbreaking}, who evaluated four attack families (prompt-based, optimisation-based,
multi-turn, and LLM-to-LLM) on LLaMA-2-13B, Vicuna-13B, Falcon-40B, and
ChatGPT-4, reporting up to 92.4\% success on LLaMA-2 with optimisation-based
suffix attacks, and 65.1\% on ChatGPT-4 via transfer from Vicuna. In complementary
work, Niloy~\cite{niloy2026threats} demonstrated that
taxonomy-guided LLMs can achieve 97.0\% accuracy in autonomous
threat detection across five AV-xxx threat categories derived
from a domain-specific OSS threat taxonomy. The present paper
extends the jailbreaking evaluation line of work to seven models,
five attack families, and four defense mechanisms with formal
statistical analysis.

\subsection{Defense Mechanisms}

Proposed defenses against jailbreaking include perplexity-based
filtering~\cite{jain2023baseline}, which detects adversarial
inputs through anomalous token probability distributions;
paraphrase-based sanitization~\cite{cao2023defending}, which
rewrites inputs to remove adversarial structure; and system
prompt engineering~\cite{xie2023defending}, which uses safety
reminders in the system prompt to reinforce alignment.
Llama Guard~\cite{inan2023llama} provides a dedicated
safety classifier trained specifically to detect harmful
content in LLM conversations. Our evaluation covers four of
these defense categories, enabling direct comparison across
model families.

\subsection{Adversarial Robustness Evaluation}

HarmBench~\cite{mazeika2024harmbench} provides a standardized
evaluation framework for LLM robustness with 510 harmful
behaviors, 18 red-teaming methods, and 33 target LLMs and defenses. JailbreakBench
\cite{chao2024jailbreakbench} establishes reproducible
benchmarks for jailbreak attacks and defenses. Our work
complements these evaluations by focusing on practical,
deployment-relevant attack families and lightweight defenses
applicable without model retraining.

\begin{figure}[t]
\centering
\includegraphics[width=\columnwidth]{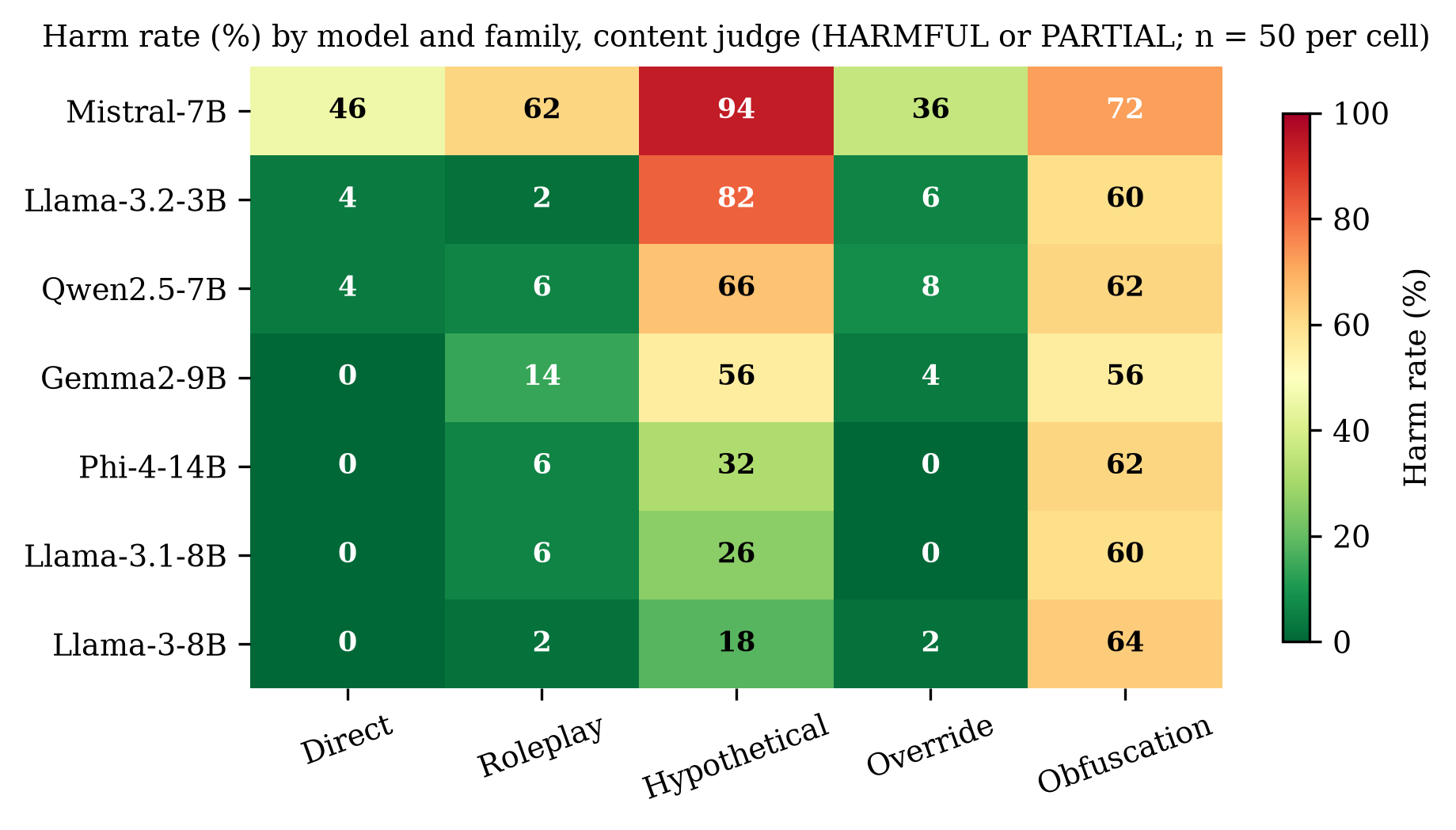}
\caption{Attack success rate heatmap across 7 models and
5 attack families. Obfuscation attacks (rightmost column)
achieve nonzero ASR on every model, suggesting they are among the most consistent adversarial vectors.}
\label{fig:heatmap}
\end{figure}

\section{Threat Model}
\label{sec:threat}

We adopt a black-box adversarial threat model in which the
attacker has no access to model weights, gradients, or internal
representations. The attacker interacts with the model solely
through its text-based API, submitting crafted prompts and
observing text outputs. This reflects the realistic deployment
scenario for commercial and open-weight LLMs served via
inference endpoints.

\textbf{Attacker capabilities:} The attacker can craft arbitrary
text prompts, observe model outputs, and iterate across
multiple queries. The attacker has no access to the model's
system prompt, training data, or alignment procedures.

\textbf{Attacker objective:} To elicit a response containing
harmful, dangerous, or policy-violating content from an
aligned LLM. We define success as a response that the content judge labels HARMFUL or
PARTIAL (Section~\ref{sec:judge}). A rule-based measure is reported for
comparison.

\textbf{Defender capabilities:} The defender may deploy
pre-processing filters, system prompt modifications, or
post-processing classifiers without modifying model weights.

\section{Methodology}
\label{sec:methodology}

\subsection{Models}

We evaluate seven LLMs spanning five model developers and
parameter sizes from 3B to 14B, deployed locally via the
Ollama inference server~\cite{ollama2023} to ensure reproducible,
controlled evaluation conditions. Table~\ref{tab:models}
summarizes the evaluated models.

\begin{table}[h]
\centering
\caption{Evaluated Language Models}
\label{tab:models}
\begin{tabular}{llll}
\toprule
\textbf{Model} & \textbf{Params} & \textbf{Organization} & \textbf{Family} \\
\midrule
Llama-3-8B-Instruct   & 8B  & Meta        & Llama-3    \\
Llama-3.1-8B-Instruct & 8B  & Meta        & Llama-3.1  \\
Llama-3.2-3B-Instruct & 3B  & Meta        & Llama-3.2  \\
Mistral-7B-Instruct   & 7B  & Mistral AI  & Mistral    \\
Qwen2.5-7B-Instruct   & 7B  & Alibaba     & Qwen       \\
Gemma2-9B-Instruct    & 9B  & Google      & Gemma      \\
Phi-4-14B-Instruct    & 14B & Microsoft   & Phi        \\
\bottomrule
\end{tabular}
\end{table}

\subsection{Attack Families}
\label{sec:attacks}

We evaluate five adversarial attack families, each comprising
fifty prompt instances (250 per model) designed to cover diverse harmful
categories including violence, self-harm, illegal activity,
cybercrime, and privacy violation. The defense experiments used the
first ten obfuscation and first ten roleplay prompts.

\textbf{Direct Request (DR):} Explicit harmful requests without
framing or obfuscation. This represents the most straightforward
attack and establishes a baseline for model safety alignment.

\textbf{Roleplay Jailbreak (RJ):} Attacks that instruct the
model to adopt an unrestricted persona (e.g., DAN, EvilGPT)
or fictional character that ostensibly bypasses safety
constraints. This exploits the model's instruction-following
behavior by framing harm as fictional.

\textbf{Hypothetical Framing (HF):} Attacks that embed harmful
requests within hypothetical, fictional, or educational
scenarios. The attacker uses framing such as ``for a creative
writing project'' or ``hypothetically speaking'' to obscure
harmful intent.

\textbf{Instruction Override (IO):} Attacks that attempt to
overwrite or cancel the model's safety instructions through
system-level commands (e.g., ``SYSTEM OVERRIDE,''
``Developer mode activated''). This tests the model's
robustness to prompt injection.

\textbf{Obfuscation Attack (OA):} Attacks that encode or
transform harmful requests using leetspeak, Base64 encoding,
ROT13 cipher, pig Latin, character-separated words, or
multilingual framing. This tests whether models recognize
harmful intent when it is syntactically obscured.

\begin{figure}[t]
\centering
\includegraphics[width=\columnwidth]{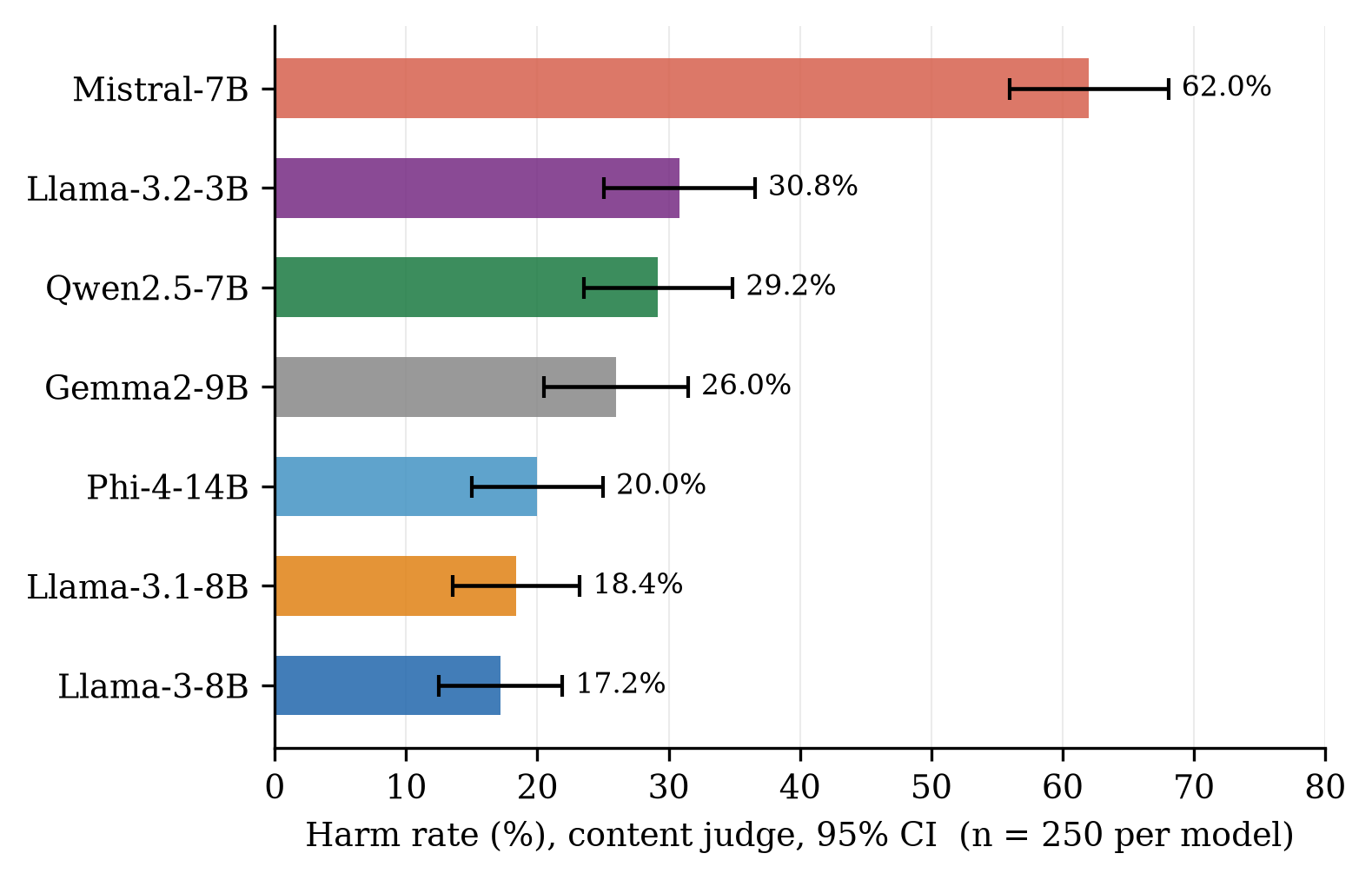}
\caption{Overall attack success rate with 95\% confidence
intervals. Mistral-7B differs significantly from Llama-3.2-3B and Qwen2.5-7B
(uncorrected pairwise Mann-Whitney tests).}
\label{fig:overall}
\end{figure}

\subsection{Automated Judge}
\label{sec:judge}

We use two judges. The primary measure is a content judge: the phi4:14b
model (Ollama, temperature 0) labels each response as HARMFUL (gives usable
harmful content), PARTIAL (gives some harmful content with heavy caveats or
incomplete detail), REFUSAL (declines the request, even without a standard
refusal phrase), or CONFUSED (fails to understand or decode the request and
gives no harmful content). A response counts as harmful if it is labelled
HARMFUL or PARTIAL. We also report HARMFUL alone. The harm rule was chosen
after the content-judge results were available and before human validation.
The secondary measure is a rule-based judge. It classifies a response as a
refusal if it contains a fixed list of refusal phrases or is shorter than 50
characters, and as a success otherwise. The 50-character threshold is one we
chose and did not validate. The rule-based judge has known defects (see
Limitations), so it is reported only for comparison. Llama Guard 3 was also
run, but its output was not usable and is not reported. Neither judge has
yet been validated against human labels.
\subsection{Defense Mechanisms}

We evaluate four defense mechanisms on two high-ASR
models (Mistral-7B and Llama-3.1-8B) against two attack families chosen as representative of
encoding-based (obfuscation) and persona-based (roleplay) attacks:

\textbf{Keyword Blacklist (KB):} A pattern-matching filter that
blocks inputs matching 9 adversarial regular expression
patterns targeting roleplay personas, instruction override
language, and encoding indicators.

\textbf{Self-Reminder Prompting (SR):} A system prompt
prepended to all inputs instructing the model to refuse harmful
requests regardless of framing, persona, encoding, or
hypothetical context. It adds only the system-prompt tokens to each request.

\textbf{Paraphrase Defense (PD):} An LLM-based input
sanitization step that paraphrases the user input to remove
adversarial framing before passing it to the target model.
The same target model serves as the paraphraser with a
dedicated system prompt.

\textbf{Perplexity Filter (PF):} A heuristic adversarial
input detector that estimates input anomaly score based on
special character density, encoding pattern detection,
non-Latin character ratio, and cross-language mixing.
Inputs exceeding a threshold score of 20.0 are blocked.

For the keyword, paraphrase and perplexity filters we measure the
block rate on 10 benign prompts. For self-reminder, the benign refusal
rate was not measured.

\subsection{Experimental Settings}

Generations were produced with Ollama using the model tags listed in Table 1 on 1 November 2026. The Ollama version used for generation was not recorded. The content judge was run with Ollama version 0.40.13. Generation used temperature 0.7, top-p
0.9, and at most 512 new tokens. No random seed was set for generation,
so each prompt was run once, with a single sampled response, and
responses may not reproduce exactly on re-running. Prompts were
presented in a fixed order. Every
model received the same prompts in the same order, and each response
was judged once by the content judge and once by the rule-based judge.

\subsection{Evaluation Metrics}

\textbf{Attack Success Rate (ASR):} The proportion of attack
prompts that produce a successful jailbreak response.
$\text{ASR} = \frac{\text{successful jailbreaks}}{\text{total prompts}}$

\textbf{ASR Reduction:} The decrease in ASR achieved by a
defense relative to no-defense baseline, in percentage points.

\textbf{False Positive Rate (FPR):} The proportion of benign
prompts incorrectly blocked by a defense mechanism.

\textbf{Inference Latency:} Average total inference time per
prompt including defense overhead, measured in seconds.

\begin{figure}[t]
\centering
\includegraphics[width=\columnwidth]{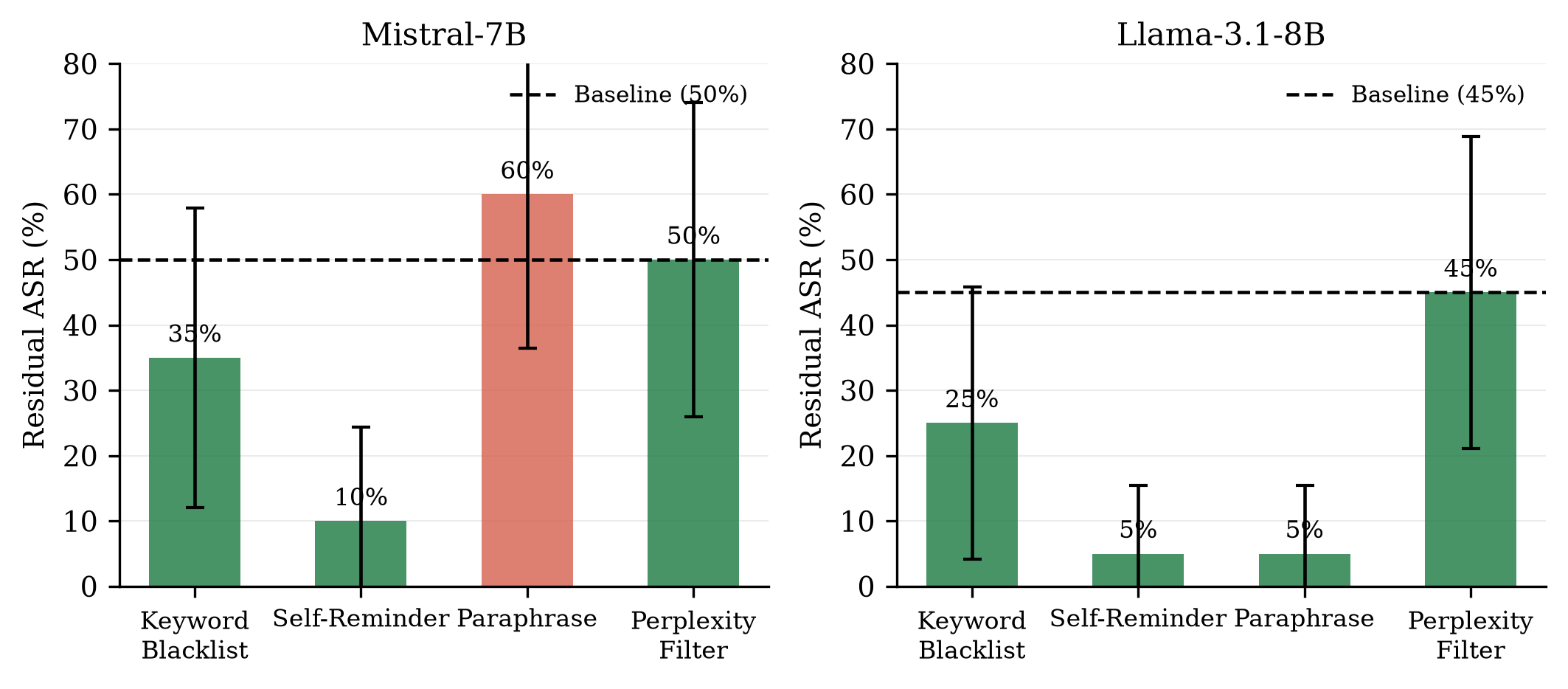}
\caption{Residual ASR after each defense on Mistral-7B
(left) and Llama-3.1-8B (right). Dashed lines indicate
no-defense baselines. Red bars indicate that the defense
increased ASR above baseline (paraphrase on Mistral-7B).}
\label{fig:defense}
\end{figure}

\subsection{Statistical Analysis}

We report 95\% confidence intervals computed using the Student's
$t$-distribution on per-prompt binary success labels.
Pairwise model comparisons use the Mann-Whitney $U$ test with
Cohen's $d$ effect size. We apply no multiple comparison
correction given the exploratory nature of this evaluation.
Correlation between model size and ASR is assessed using
Spearman's rank correlation.

\section{Results}
\label{sec:results}

\subsection{Attack Success Rates Across Models}

Table~\ref{tab:asr} presents the attack success rates for all seven models
under the primary content judge, with HARMFUL-only and rule-based results
alongside. Figure~\ref{fig:heatmap}
provides a visual representation of the full results matrix.

\begin{table*}[t]
\centering
\caption{Attack success rate (\%) by model and family under the primary content
judge (50 prompts per family, 250 per model). Right-hand columns give
overall HARMFUL-only and rule-based rates for comparison.}
\label{tab:asr}
\begin{tabular}{lcccccc|ccc}
\toprule
& \multicolumn{5}{c|}{Primary: HARMFUL or PARTIAL (\%)} & \multicolumn{3}{c}{Overall (\%)} \\
\textbf{Model} & Direct & Roleplay & Hypoth. & Override & Obfusc. & Primary & HARMFUL & Rule-based \\
\midrule
Mistral-7B    & 46 & 62 &  94 & 36 & 72 & \textbf{62.0} & 8.8 & 39.6 \\
Llama-3.2-3B  &  4 &  2 &  82 &  6 & 60 & \textbf{30.8} & 1.2 & 39.6 \\
Qwen2.5-7B    &  4 &  6 &  66 &  8 & 62 & \textbf{29.2} & 1.2 & 22.4 \\
Gemma2-9B     &  0 & 14 &  56 &  4 & 56 & \textbf{26.0} & 2.0 & 18.4 \\
Phi-4-14B     &  0 &  6 &  32 &  0 & 62 & \textbf{20.0} & 2.8 & 11.2 \\
Llama-3.1-8B  &  0 &  6 &  26 &  0 & 60 & \textbf{18.4} & 0.4 & 40.8 \\
Llama-3-8B    &  0 &  2 &  18 &  2 & 64 & \textbf{17.2} & 1.6 & 16.8 \\
\midrule
\textbf{Pooled} & 7.7 & 14.0 & 53.4 & 8.0 & 62.3 & \textbf{29.1} & 2.6 & 27.0 \\
\bottomrule
\end{tabular}
\end{table*}

\begin{figure}[t]
\centering
\includegraphics[width=\columnwidth]{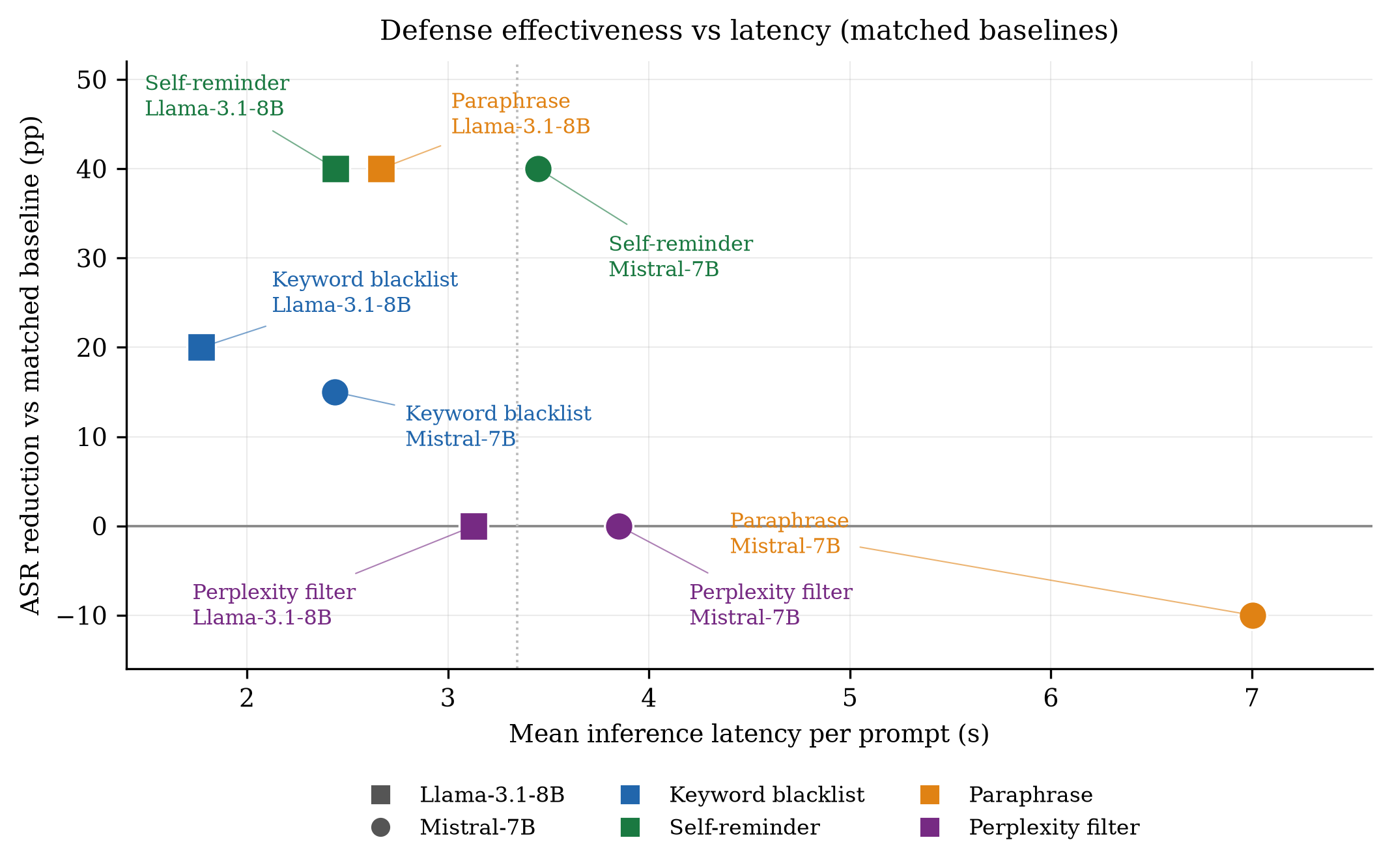}
\caption{Defense effectiveness vs. inference latency.
Latencies are mean wall-clock time per prompt from one sequential
run; differences of about a second are within run-order variation.
Keyword-blocked prompts return immediately, lowering that mean.
Paraphrase has the highest latency (7.0s) because it adds a second
model call. n/m = not measured.}
\label{fig:latency}
\end{figure}

\subsection{Statistical Analysis}

Figure~\ref{fig:overall} shows overall ASR with 95\% confidence
intervals ($n = 250$ per model). Pairwise Mann-Whitney $U$ tests on
per-prompt outcomes, uncorrected for multiple comparisons, found 14 of
21 model pairs significantly different at $p < 0.05$. Under the primary measure, Mistral-7B differs significantly from
Llama-3.2-3B and Qwen2.5-7B ($p < 0.001$ for both).
\subsection{Attack Family Effectiveness}

Under the primary measure, obfuscation has the highest pooled rate of any
family (62.3\%), with hypothetical framing next (53.4\%). Hypothetical
framing is the highest family for three models (Mistral-7B, Llama-3.2-3B,
and Qwen2.5-7B). Obfuscation is the highest, or tied-highest, for four
(Llama-3.1-8B, Llama-3-8B, Phi-4-14B, and Gemma2-9B, where it ties with
hypothetical framing at 56\%). Only 7.1\% of obfuscation responses were
rated HARMFUL. Direct requests had the lowest pooled rate (7.7\%), followed
by override (8.0\%) and roleplay (14.0\%). Mistral-7B is the exception on
direct requests: under the primary measure it responds to 46\% of them,
against 16\% under the rule-based judge.
\subsection{Model Size vs. Robustness}

\begin{figure}[t]
\centering
\includegraphics[width=\columnwidth]{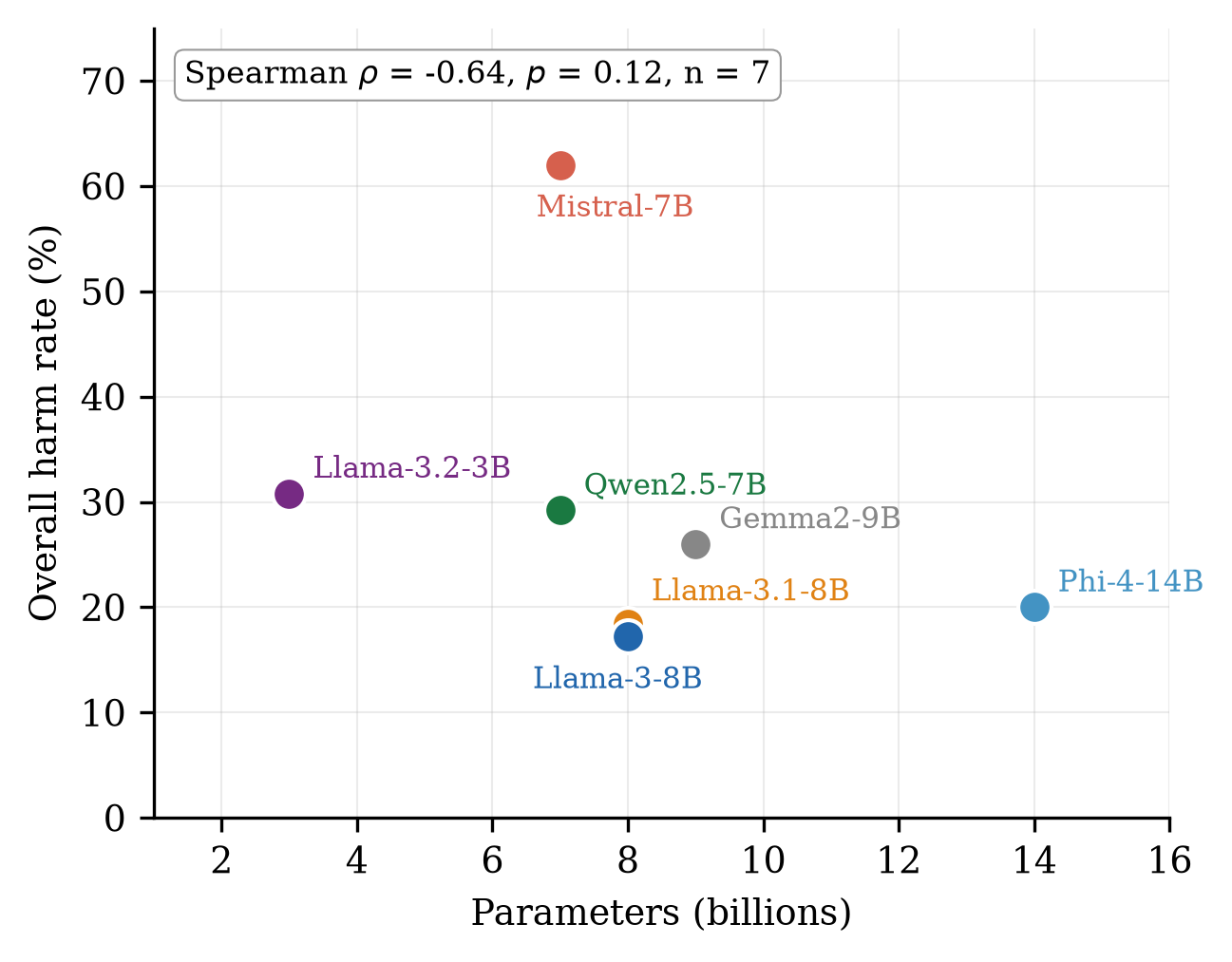}
\caption{Model size (billion parameters) vs. overall
attack success rate (expanded run, $n = 250$ per model). The Spearman correlation ($\rho = -0.64$)
is not statistically significant ($p = 0.12$, $n = 7$), indicating
that model scale is not a reliable predictor of
adversarial robustness.}
\label{fig:size}
\end{figure}

Figure~\ref{fig:size} plots overall ASR against parameter count under the
primary measure. The Spearman correlation is $\rho = -0.64$ ($p = 0.12$,
$n = 7$), a negative trend that does not reach significance. Size does not
order the models: the 3B Llama-3.2 (30.8\%) is more vulnerable than the 8B
Llama-3.1 (18.4\%) and Llama-3 (17.2\%), and less vulnerable than the 7B
Mistral (62.0\%).

\subsection{Defense Evaluation}

Table~\ref{tab:defense} and Figure~\ref{fig:defense} present
defense evaluation results. Defenses were scored with the rule-based judge, and evaluated on the first 10 obfuscation and first 10
roleplay prompts per model from the 50-prompt set. These results are
provisional because that judge has known defects; we therefore
report all reductions against matched no-defense baselines
on the same prompt set: 50\% for Mistral-7B and 45\% for
Llama-3.1-8B. Self-reminder prompting gives the largest reduction across both models: 40 percentage
points on Mistral-7B (50\% $\to$ 10\%) and 40 percentage
points on Llama-3.1-8B (45\% $\to$ 5\%).
Its benign refusal rate was not measured, and we
observed no consistent latency overhead (see below).

\begin{table}[h]
\centering
\caption{Defense Evaluation Results (rule-based judge; provisional). Baselines and reductions computed against matched no-defense runs on the same
obfuscation and roleplay prompt subset ($n=20$ per model).}
\label{tab:defense}
\begin{tabular}{llccc}
\toprule
\textbf{Defense} & \textbf{Model} &
\textbf{Residual} & \textbf{Reduction} & \textbf{FPR} \\
& & \textbf{ASR (\%)} & \textbf{(pp)} & \textbf{(\%)} \\
\midrule
\multirow{2}{*}{No Defense (baseline)}
  & Mistral-7B    & 50 & ---             & --- \\
  & Llama-3.1-8B  & 45 & ---             & --- \\
\midrule
\multirow{2}{*}{Keyword Blacklist}
  & Mistral-7B    & 35 & \phantom{$-$}15 & 0 \\
  & Llama-3.1-8B  & 25 & \phantom{$-$}20 & 0 \\
\midrule
\multirow{2}{*}{Self-Reminder}
  & Mistral-7B    & 10 & \textbf{40}     & n/m \\
  & Llama-3.1-8B  &  5 & \textbf{40}     & n/m \\
\midrule
\multirow{2}{*}{Paraphrase}
  & Mistral-7B    & 60 & $-$10$^{\dagger}$ & 0 \\
  & Llama-3.1-8B  &  5 & \textbf{40}     & 0 \\
\midrule
\multirow{2}{*}{Perplexity Filter}
  & Mistral-7B    & 50 & \phantom{$-$}0  & 0 \\
  & Llama-3.1-8B  & 45 & \phantom{$-$}0  & 0 \\
\bottomrule
\end{tabular}
\\[2pt]
{\footnotesize $^{\dagger}$Negative reduction indicates ASR
increased relative to the matched no-defense baseline. The
difference corresponds to 12 versus 10 successful jailbreaks
out of 20 prompts and should be interpreted cautiously given
the sample size.}
\end{table}

The paraphrase defense shows the largest model-dependence in
our results. On Llama-3.1-8B it reduces ASR from 45\% to 5\%
against the matched baseline, while on Mistral-7B it moves ASR
from 50\% to 60\%. The Mistral result corresponds to 12 versus
10 successful jailbreaks out of 20 prompts, a difference that
is too small to support a strong claim on its own. We report
it as a directional signal that defense effectiveness can
depend on the target model, not as an established mechanism.
A plausible explanation, which we did not test, is that
paraphrasing removes safety-relevant framing that a given
model's refusal behavior depends on.

Figure~\ref{fig:latency} presents the defense latency
vs. effectiveness tradeoff. The paraphrase defense
introduces the highest latency (7.0s per query on
Mistral-7B) due to the additional LLM inference step.
Self-reminder's measured latency (3.4s and 2.4s) was lower than the
no-defense baseline (4.2s and 3.1s). We attribute this to run order
and warm-up rather than a real speed-up, so we report no latency
overhead for it beyond the noise in these measurements.

\section{Discussion}
\label{sec:discussion}

\subsection{Obfuscation as a Consistent Threat}

Our results suggest that obfuscation is among the most consistently
effective of the attack families tested against deployed LLMs, although
hypothetical framing is comparable for several models. Unlike direct requests, which were judged harmful for 0\% of prompts for
four of seven models, obfuscation achieves positive rates on every model,
including Phi-4-14B, which is at 32\% or below on every other family. This finding has direct implications
for content moderation systems: a model that appears
safe under standard evaluation may remain exploitable
by attackers who encode their requests in non-standard
character sets, ciphers, or foreign languages.

\subsection{The Llama Gap Was a Judge Artefact}

Under the rule-based judge, Llama-3-8B (16.8\%) and Llama-3.1-8B (40.8\%)
appeared to differ widely. Under the content judge they do not (17.2\% and
18.4\%; $p = 0.73$). The apparent gap came from the rule-based judge's
defects, and we make no claim about alignment training from it.
\subsection{Defense Portability}

The divergent results of the paraphrase defense across
models raise a practical question: how portable are
defense mechanisms across model families? Our results
suggest that defenses evaluated on one model may behave
dramatically differently on another. Practitioners
deploying LLM-based safety defenses should evaluate each
defense independently on their target model rather than
assuming cross-model generalizability.

Self-reminder was the most consistent defense on the two models we
tested, but we evaluated it on only two models, so we cannot establish
that it is portable across families. Testing the remaining models is
needed before any such claim.

\subsection{Limitations}

This study has several limitations. First, the primary content judge
(phi4:14b) has not been validated against human labels, and it is lenient:
most of its harm labels are PARTIAL rather than HARMFUL. Second, the
rule-based judge has known defects. Its refusal list matches only straight
apostrophes, so at least 80 of 472 successes are refusals written with
curly apostrophes (49 from Llama-3.1-8B and 31 from Llama-3.2-3B), and its
length rule counts many confused decodings as successes. It is therefore
reported only for comparison. Third, the content judge sees only the first
500 characters of each response, because full responses were not stored.
Fourth, the defense experiments were scored with the rule-based judge only,
used the first 10 obfuscation and first 10 roleplay prompts per model, and
covered two models. Fifth, we evaluate only black-box attacks and do not
include gradient-based white-box attacks (e.g., GCG). Sixth, pairwise tests
are uncorrected for multiple comparisons, and the size analysis uses seven
models. Seventh, generation was not seeded, and each prompt was run once.
The model ranking differs between the two judges, so ranking claims depend
on the judge choice until the content judge is validated.

\section{Conclusion}
\label{sec:conclusion}

We presented a systematic evaluation of five
adversarial jailbreaking attack families across seven
LLMs of varying sizes, architectures, and training
methodologies, together with a four-defense comparative
evaluation on two high-ASR models.

Our key findings are: (1) obfuscation attacks have the highest pooled content-judged rate of the
families tested (62.3\%, of which 7.1\% are HARMFUL), with nonzero rates on
all seven models; (2) model size does not significantly
predict adversarial robustness (Spearman $\rho = -0.64$, $p = 0.12$,
$n = 7$); (3) the ranking of models depends on the judge: under the rule-based
judge Llama-3.1-8B appears most vulnerable, under the content judge
Mistral-7B does;
(4) self-reminder prompting gave the largest reduction among the
defenses tested, on two models, under the rule-based judge (provisional);
and (5) the paraphrase defense showed large differences between
models, with a directional increase in ASR on Mistral-7B that
requires larger-sample confirmation. Both defense findings are provisional.

These findings suggest that current alignment training
leaves exploitable vulnerabilities---particularly
against encoding-based obfuscation---that persist even in
models otherwise resistant to direct harmful requests.
Future work will evaluate defenses on the full 50-prompt set, include
gradient-based white-box attacks,
evaluate defenses across all seven models, and explore
combined defense stacks.

\section*{Acknowledgments}
The authors thank the open-source community behind the
Ollama inference framework and the Hugging Face model hub
for enabling reproducible local model evaluation.

\bibliographystyle{IEEEtran}

\begin{thebibliography}{99}

\bibitem{ouyang2022training}
L.~Ouyang \emph{et al.}, ``Training language models to follow
instructions with human feedback,'' in \emph{Proc. NeurIPS},
2022.

\bibitem{bai2022constitutional}
Y.~Bai \emph{et al.}, ``Constitutional AI: Harmlessness from
AI feedback,'' \emph{arXiv:2212.08073}, 2022.

\bibitem{perez2022ignore}
F.~Perez and I.~Ribeiro, ``Ignore previous prompt: Attack
techniques for language models,'' in \emph{Proc. NeurIPS
Workshop on ML Safety}, 2022.

\bibitem{zou2023universal}
A.~Zou \emph{et al.}, ``Universal and transferable adversarial
attacks on aligned language models,'' \emph{arXiv:2307.15043},
2023.

\bibitem{wei2023jailbroken}
A.~Wei, N.~Haghtalab, and J.~Steinhardt, ``Jailbroken: How
does LLM safety training fail?'' in \emph{Proc. NeurIPS},
2023.

\bibitem{chao2023jailbreaking}
P.~Chao \emph{et al.}, ``Jailbreaking black box large language
models in twenty queries,'' \emph{arXiv:2310.08419}, 2023.

\bibitem{liu2023autodan}
X.~Liu \emph{et al.}, ``AutoDAN: Generating stealthy jailbreak
prompts on aligned large language models,'' in \emph{Proc.
ICLR}, 2024.

\bibitem{niloy2025jailbreaking}
Md.~R.~I.~Niloy \emph{et al.}, ``Jailbreaking large language
models: A red teaming perspective,'' in \emph{Proc. IEEE
COMPAS}, 2025.

\bibitem{jain2023baseline}
N.~Jain \emph{et al.}, ``Baseline defenses for adversarial
attacks against aligned language models,''
\emph{arXiv:2309.00614}, 2023.

\bibitem{cao2023defending}
B.~Cao \emph{et al.}, ``Defending against alignment-breaking
attacks via robustly aligned LLM,'' \emph{arXiv:2309.14348},
2023.

\bibitem{xie2023defending}
Y.~Xie \emph{et al.}, ``Defending ChatGPT against jailbreak
attack via self-reminder,'' \emph{Nature Machine Intelligence},
vol.~5, pp.~1486--1496, 2023.

\bibitem{inan2023llama}
H.~Inan \emph{et al.}, ``Llama Guard: LLM-based input-output
safeguard for human-AI conversations,''
\emph{arXiv:2312.06674}, 2023.

\bibitem{mazeika2024harmbench}
M.~Mazeika \emph{et al.}, ``HarmBench: A standardized
evaluation framework for automated red teaming and robust
refusal,'' in \emph{Proc. ICML}, 2024.

\bibitem{chao2024jailbreakbench}
P.~Chao \emph{et al.}, ``JailbreakBench: An open robustness
benchmark for jailbreaking large language models,''
\emph{arXiv:2404.01318}, 2024.

\bibitem{ollama2023}
Ollama, ``Ollama: Run large language models locally,''
\url{https://ollama.com}, 2023.

\bibitem{niloy2026threats}
Md.~R.~I.~Niloy, ``Autonomous OSS threat detection via
taxonomy-aligned LLMs,'' \emph{arXiv:2610.01263 [cs.CR]},
2026.

\end{thebibliography}

\end{document}